\documentclass[aps,prl,twocolumn,showpacs,preprintnumbers]{revtex4}

\usepackage{graphicx}

\let\bm=\bibitem

\def\ft#1#2{{\textstyle{{\scriptstyle #1}\over {\scriptstyle #2}}}}
\def\fft#1#2{{#1 \over #2}}

\def\del{\partial}
\def\nn{\nonumber}
\def\sst#1{{\scriptscriptstyle #1}}
\def\0{{\sst{(0)}}}
\def\1{{\sst{(1)}}}
\def\2{{\sst{(2)}}}
\def\3{{\sst{(3)}}}
\def\4{{\sst{(4)}}}
\def\5{{\sst{(5)}}}
\def\6{{\sst{(6)}}}
\def\7{{\sst{(7)}}}
\def\8{{\sst{(8)}}}
\def\im{{\rm i}}

\newcommand{\bea}{\begin{eqnarray}}
\newcommand{\eea}{\end{eqnarray}}
\newcommand{\be}{\begin{equation}}
\newcommand{\ee}{\end{equation}}

\begin{document}

\preprint{UPR1116-T\ \ \ \ MIFP-05-09\ \ \ \ 
Alberta Thy-06-05\ \ \ \ {\bf hep-th/0504225}}

\title{New Einstein-Sasaki Spaces in Five and Higher Dimensions}

\author{M. Cveti\v c,$^1$ H. L\"u,$^{2}$ 
Don N. Page,$^{3}$ C.N. Pope$^{2}$}
\affiliation{%
${}^1\!\!\!$ Department of Physics and Astronomy, University
of Pennsylvania, Philadelphia, PA 19104, USA
\\
${}^2\!\!\!$ George P. \& Cynthia W.
Mitchell Institute for Fundamental Physics,
Texas A\&M University, College Station, TX 77843, USA\\
${}^3\!\!\!$ Theoretical Physics Institute, 412 Physics Laboratory,
University of Alberta, Edmonton, Canada
}%

\date{April 27, 2005}

\begin{abstract}

    We obtain infinite classes of new Einstein-Sasaki metrics on
complete and non-singular manifolds.  They arise, after
Euclideanisation, from BPS limits of the rotating Kerr-de Sitter black
hole metrics.  The new Einstein-Sasaki spaces $L^{p,q,r}$ in five
dimensions have cohomogeneity 2, and $U(1)\times U(1)\times U(1)$
isometry group.  They are topologically $S^2\times S^3$. 
Their AdS/CFT duals will describe quiver theories on
the four-dimensional boundary of AdS$_5$.  We also obtain new
Einstein-Sasaki spaces of cohomogeneity $n$ in all odd dimensions
$D=2n+1\ge 5$, with $U(1)^{n+1}$ isometry.

\end{abstract}

\pacs{04.20.Jb, 04.50.+h}
\maketitle

   The AdS/CFT correspondence \cite{mal,guklpo} relates bulk solutions
in five-dimensional gauged supergravities to conformal field theories
on the four-dimensional boundary.  The gauged supergravities arise
through dimensional reduction of type IIB string theory on compact
five-dimensional Einstein spaces of positive Ricci curvature.  In
order to obtain supersymmetry in the reduced five-dimensional theory,
it is necessary that the compact five-dimensional Einstein space $K_5$
admit a Killing spinor; {\it i.e.}, that it be an Einstein-Sasaki
space.

   The most studied case is when $K_5$ is taken to be the 5-sphere,
which admits the maximal number, 4, of Killing spinors.  Upon
reduction one obtains a five-dimensional supergravity with ${\cal
N}=8$ supersymmetry and $SO(6)$ gauge fields.  The corresponding
boundary theory is an ${\cal N}=4$ supersymmetric superconformal field
theory.  Another extensively studied case is when $K_5$ is $T^{1,1}$,
which is a homogeneous Einstein-Sasaki space with $SU(2)^2\times U(1)$
isometry.  Recently, an infinite class of five-dimensional
Einstein-Sasaki spaces was obtained \cite{gamaspwa}, all of which can
provide new five-dimensional supergravities, and hence field-theory
duals on the four-dimensional boundary.  These Einstein-Sasaki spaces,
denoted by $Y^{p,q}$, are characterised by the two coprime positive
integers $p$ and $q$ with $q<p$.  In the construction in
\cite{gamaspwa}, a local family of Einstein-Sasaki metrics with a
non-trivial continuous parameter was first obtained, and then it was
shown that if the parameter takes rational values $p/q$ in the
appropriate range, the metrics extend smoothly onto the complete and
non-singular manifolds $Y^{p,q}$.  The spaces have $SU(2)\times
U(1)\times U(1)$ isometry.

    It was subsequently shown \cite{hasaya} that the Einstein-Sasaki
spaces $Y^{p,q}$ could be obtained in a straightforward manner by
taking a certain limit of the Euclideanised five-dimensional
Kerr-de Sitter black hole metrics found in \cite{hawhuntay}.
Specifically, after Euclideanisation the two rotation parameters $a$
and $b$ were set equal, and allowed to approach the limiting value
that corresponds, in the Lorentzian regime, to having rotation at the
speed of light at infinity.

    In this paper, we construct a vastly greater number of
Einstein-Sasaki spaces, in which a similar limit is taken but without
requiring the rotation parameters to be equal.  By this means, we
first obtain a family of five-dimensional local Einstein-Sasaki
metrics with two non-trivial continuous parameters.  We then show that
if these are appropriately restricted to be rational, the metrics
extend smoothly onto complete and non-singular manifolds, which we
denote by $L^{p,q,r}$, where $p$, $q$ and $r$ are coprime positive
integers with $0<p\le q$, $0<r<p+q$, and with $p$ and $q$ each coprime to $r$ 
and to $s=p+q-r$. The metrics have $U(1)\times
U(1)\times U(1)$ isometry in general, enlarging to $SU(2)\times
U(1)\times U(1)$ in the special case $p+q=2r$, which reduces to the
previously-obtained spaces $Y^{p,q}=L^{p-q,p+q,p}$.  The new
Einstein-Sasaki spaces $L^{p,q,r}$ provide backgrounds for dual quiver
field theories on the four-dimensional boundary of the corresponding
five-dimensional gauged supergravity.

   The local Einstein-Sasaki metrics that we shall construct are
obtained from the rotating AdS black hole metrics in $D=5$ dimensions
\cite{hawhuntay} and in $D>5$ \cite{gilupapo1,gilupapo2}.  Our
principal focus will be on the Euclidean-signature case with positive
Ricci curvature, but it is helpful to think first of the metrics in
the Lorentzian regime, with negative cosmological constant
$\lambda=-g^2$. It was shown in \cite{gibperpop} that the energy and
angular momenta of the $D=2n+1$ dimensional Kerr-AdS black holes are
given by
\be
E = \fft{m\, {\cal A}_{D-2}}{4\pi(\prod_j \Xi_j)}\Big(\sum_{i=1}^n 
        \fft1{\Xi_i} -\ft12\Big)\,,\quad
J_i = \fft{m a_i\, {\cal A}_{D-2}}{4\pi \Xi_i (\prod_j \Xi_j)}\,,
\label{ejrels}
\ee
where ${\cal A}_{D-2}$ is the volume of the unit $(D-2)$-sphere,
$\Xi_i= 1- g^2 a_i^2$ and $a_i$ are the $n$ independent rotation
parameters.  As discussed in \cite{cvgilupo}, the BPS limit can be
found by studying the eigenvalues of the Bogomol'nyi matrix arising in
the AdS superalgebra from the anticommutator of the supercharges.  In
$D=5$, these eigenvalues are then proportional to $E \pm g J_1 \pm g
J_2$.  The BPS limit is achieved when one or more of the eigenvalues
vanishes.  For just one zero eigenvalue, the four cases are equivalent
under reversals of the angular velocities, so we may without loss of
generality consider $E-g J_1 -g J_2=0$.  From (\ref{ejrels}), we see
that this is achieved by taking a limit in which $g a_1$ and $g a_2$
tend to unity, namely, by setting $g a_1=1-\ft12\epsilon \alpha$,
$ga_2=1-\ft12\epsilon\beta$, rescaling $m$ according to
$m=m_0\epsilon^3$, and sending $\epsilon$ to zero.  As we shall see,
the metric remains non-trivial in this limit.  An equivalent
discussion in the Euclidean regime leads to the conclusion that in the
corresponding limit, one obtains five-dimensional Einstein metrics
admitting a Killing spinor. [The above scaling limit in the Lorentzian
regime, for the special case $\alpha=\beta$, was studied recently in
\cite{cvgasi}.]

   To present our new Einstein-Sasaki metrics, we start with the
five-dimensional rotating AdS black hole solutions, and Euclideanise
by making the analytic continuations $t\rightarrow \im \sqrt\lambda\,
\tau\,, \quad \ell\rightarrow \im \sqrt{\lambda}\,,\quad a\rightarrow
\im a\,,\quad b\rightarrow \im b $ in the metric (5.22) of
\cite{hawhuntay}. Next, we implement the ``BPS scaling limit,'' by
setting
\bea
&&a=\lambda^{-\ft12} (1 - \ft12\alpha\,\epsilon)\,,\quad
b=\lambda^{-\ft12} (1 - \ft12\beta\,\epsilon)\,,\nn\\
&&r^2=\lambda^{-1} (1 - x\epsilon)\,,\quad
M=\ft12\lambda^{-1} \mu \epsilon^3
\eea
and then sending $\epsilon\rightarrow 0$.  The metric becomes
\be
\lambda\,ds_5^2 = (d\tau + \sigma)^2 + ds_4^2\,,\label{5met}
\ee
where
\bea
ds_4^2 &=& \fft{\rho^2\,dx^2}{4\Delta_x} +
\fft{\rho^2\,d\theta^2}{\Delta_\theta} +
\fft{\Delta_x}{\rho^2} (\fft{\sin^2\theta}{\alpha} d\phi +
\fft{\cos^2\theta}{\beta} d\psi)^2\nn\\
&& + \fft{\Delta_\theta\sin^2\theta\cos^2\theta}{\rho^2} 
  (\fft{\alpha - x}{\alpha}
d\phi - \fft{\beta - x}{\beta} d\psi)^2\,,\nn\\
\sigma &=& \fft{(\alpha -x)\sin^2\theta}{\alpha} d\phi +
\fft{(\beta-x)\cos^2\theta}{\beta} d\psi\,,\label{d4met}\\
\Delta_x &=& x (\alpha -x) (\beta - x) - \mu\,,\quad
    \rho^2=\Delta_\theta-x\,,\nn\\
\Delta_\theta &=& \alpha\, \cos^2\theta + \beta\, \sin^2\theta
\nn
\eea
It is eay to check that the four-dimensional metric in
(\ref{d4met}) is Einstein.  The parameter $\mu$ is trivial,
and can be set to any non-zero constant, say $\mu=1$, by rescaling
$\alpha$, $\beta$ and $x$.  The metrics depend on two non-trivial
parameters, which we can take to be $\alpha$ and $\beta$ at fixed
$\mu$.  It is sometimes convenient to retain $\mu$, allowing
it to be determined as the product of the three roots $x_i$ of
$\Delta_x$.

   The five-dimensional metric can be viewed as $U(1)$ bundle over a
four-dimensional Einstein-K\"ahler metric, with K\"ahler 2-form given
by $J=\ft12 d\sigma$.  It is straightforward to verify that $J$ indeed
gives an almost complex structure tensor, and that it is covariantly
constant.  This demonstrates that the $D=4$ metric is
Einstein-K\"ahler and hence the $D=5$ metric is Einstein-Sasaki, with
$R_{\mu\nu}= 4\lambda g_{\mu\nu}$.

   Having obtained the local form of the five-dimensional
Einstein-Sasaki metrics, we can now turn to an analysis of the global
structure.  The metrics are in general of cohomogeneity 2, with toric
principal orbits $U(1)\times U(1)\times U(1)$.  The orbits degenerate
at $\theta=0$ and $\theta=\ft12 \pi$, and at the roots of the cubic
function $\Delta_x$ appearing in (\ref{d4met}).  In order to obtain
metrics on complete non-singular manifolds, one must impose
appropriate conditions to ensure that the collapsing orbits extend
smoothly, without conical singularities, onto the degenerate surfaces.
If this is achieved, one can obtain a metric on a non-singular
manifold, with $0\le\theta\le\ft12\pi$ and $x_1\le x\le x_2$, where
$x_1$ and $x_2$ are two adjacent real roots of $\Delta_x$. In fact,
since $\Delta_x$ is negative at large negative $x$ and positive at
large positive $x$, and since we must also have $\Delta_x>0$ in the
interval $x_1<x<x_2$, it follows that $x_1$ and $x_2$ must be the
smallest two roots of $\Delta_x$.

   The easiest way to analyse the behaviour at each collapsing orbit
is to examine the associated Killing vector $\ell$ whose length
vanishes at the degeneration surface. By normalising the Killing vector
so that its ``surface gravity'' $\kappa$ is equal to unity, one
obtains a translation generator $\del/\del \chi$ where $\chi$ is a
local coordinate near the degeneration surface, and the metric extends
smoothly onto the surface if $\chi$ has period $2\pi$.  The ``surface
gravity'' is
\be
\kappa^2 = \fft{g^{\mu\nu}\, (\del_\mu \ell^2)(\del_\nu 
\ell^2)}{4\ell^2}
\ee
in the limit that the degeneration surface is reached.

   The normalised Killing vectors that vanish at the degeneration
surfaces $\theta=0$ and $\theta=\ft12\pi$ are simply given by $\del/\del
\phi$ and $\del/\del\psi$ respectively.  At the degeneration surfaces
$x=x_1$ and $x=x_2$, we find that the associated normalised Killing
vectors $\ell_1$ and $\ell_2$ are given by
\be
\ell_i = c_i\, \fft{\del}{\del \tau} + a_i\, \fft{\del}{\del\phi} +
            b_i\, \fft{\del}{\del\psi}\,,\label{ells}
\ee
where the constants $c_i$, $a_i$ and $b_i$ are given by
\bea
a_i &=& \fft{\alpha c_i}{x_i - \alpha}\,,\qquad
b_i = \fft{\beta c_i}{x_i-\beta}\,,\nn\\
c_i &=& \fft{(\alpha-x_i)(\beta-x_i)}{2(\alpha+\beta) x_i - 
\alpha\beta - 3 x_i^2}\,.\label{abci}
\eea

   Since we have a total of four Killing vectors $\del/\del\phi$,
$\del/\del\psi$, $\ell_1$ and $\ell_2$ that span a three-dimensional
space, there must exist a linear relation amongst
them.  Since they all generate translations with a $2\pi$ period repeat, 
it follows that unless the coefficients in the linear
relation are rationally related, then by taking integer combinations
of translations around the $2\pi$ circles, one could generate a
translation implying an identification of arbitrarily nearby points in
the manifold.  Thus one has the requirement for obtaining a
non-singular manifold that the linear relation between the four
Killing vectors must be expressible as
\be
p \ell_1 + q \ell_2 + r\, \fft{\del}{\del\phi} + s \, \fft{\del}{\del\psi}=0
\label{lincomb}
\ee
for {\it integer} coefficients $(p,q,r,s)$, which may be assumed to be
coprime.  All subsets of three of the four integers must be coprime
too, since if any three had a common divisor $k$, then dividing
(\ref{lincomb}) by $k$ would show that the direction associated with
the Killing vector whose coefficient was not divisible by $k$ would be
identified with period $2\pi/k$, thus leading to a conical
singularity.  Furthermore, $p$ and $q$ must each be coprime to each of
$r$ and $s$, since otherwise at the surfaces where $\theta=0$ or
$\ft12\pi$ {\it and} $x=x_1$ or $x=x_2$ -- at which one of
$\del/\del\phi$ or $\del/\del/\psi$ {\it and} simulataneously one of
$\ell_1$ or $\ell_2$ vanish -- there would be conical singularities.
(We are grateful to J. Sparks for pointing this out to us; see
hep-th/0505027, hep-th/0505211, hep-th/0505220.)

   From (\ref{lincomb}), and (\ref{ells}), we have
\bea
&& p a_1 + q a_2 + r=0\,,\qquad
p b_1 + q b_2 + s=0\,,\nn\\
&& p c_1 + q c_2=0\,.\label{klmn}
\eea
It then follows that all ratios
between the four quantities
\be
a_1 c_2-a_2 c_1\,,\quad
 b_1 c_2 -b_2 c_1 \,,\quad
c_1\,,\quad c_2\label{rationals}
\ee
must be rational.  Thus to obtain a metric that extends smoothly
onto a complete and non-singular manifold, we must choose the parameters
in (\ref{d4met}) so that the rationality of the ratios is achieved.  In
fact it follows from (\ref{abci}) that
\be
1+a_i + b_i + 3 c_i = 0\,,\label{abcid}
\ee
for all roots $x_i$, and using this one can show that there are only
two independent rationality conditions following from the requirements
of rational ratios for the four quantities in (\ref{rationals}).  One
can also see from (\ref{abcid}) that
\be
p+q-r-s=0\,,\label{klmnrel}
\ee
so the further requirement that all triples among the 
$(p,q,r,s)$ also be coprime is automatically satisfied. 

   The upshot from the above discussion is that we can have complete and
non-singular five-dimensional Einstein-Sasaki spaces $L^{p,q,r}$,
where
\be
pc_1+q c_2=0\,,\quad pa_1 + q a_2 + r=0\,.
\ee
These equations and (\ref{abcid}) allow one to solve for $\alpha$,
$\beta$ and the roots $x_1$ and $x_2$, for positive coprime integer
triples $(p,q,r)$.  The requirements $0\le x_1\le x_2 \le x_3$, and
$\alpha\ge x_2$, $\beta\ge x_2$, restrict the integers to the domain
$0< p \le q$ and $0 <r < p+q$. All such coprime triples with $p$ and $q$ each 
coprime to $r$ and $s$ yield complete and
non-singular Einstein-Sasaki spaces $L^{p,q,r}$, and so we get infinitely
many new examples.

   The volume of $L^{p,q,r}$ (with $\lambda=1$) is given by 
\be
V=\fft{\pi^2(x_2-x_1)
(\alpha+\beta -x_1-x_2)\Delta\tau}{2k\alpha\beta}\,,
\ee
where $\Delta\tau$ is the period of the coordinate $\tau$, and
$k=\hbox{gcd}\, (p,q)$. Note that the $(\phi,\psi)$ torus is factored
by a freely-acting $Z_k$, along the diagonal.  $\Delta\tau$ is given
by the minimum repeat distance of $2\pi c_1$ and $2\pi c_2$, and so
$\Delta\tau=2\pi k |c_1|/q$. There is a quartic equation expressing
$V$ purely in terms of $(p,q,r)$.  Writing $V=\pi^3 (p+q)^3
W/(8pqrs)$, we find
\bea
0&=&(1-f^2)(1-g^2)h_-^4 + 2h_-^2[2(2-h_+)^2-3h_-^2]W\nn\\
&&+ [8h_+(2-h_+)^2-h_-^2(30+9h_+)]W^2 \nn\\
&&+8(2-9h_+) W^3 - 27 W^4\label{quartic}
\eea
where $f=(q-p)/(p+q)$, $g=(r-s)/(p+q)$, and $h_\pm=f^2\pm g^2$.  The
central charge of the dual field theory is rational if $W$ is rational,
which is easily achieved.

   If one sets $p+q=2r$, {\it i.e.}, $r=s$, implying $\alpha$ and
$\beta$ become equal, our Einstein-Sasaki metrics reduce to those in
\cite{gamaspwa}, and the conditions we have discussed for achieving
complete non-singular manifolds reduce to the conditions for the
$Y^{p,q}$ obtained there, with $Y^{p,q}= L^{p-q,p+q,p}$.  The quartic
(\ref{quartic}) then factorises to quadratics with rational
coefficients, giving the volumes found in \cite{gamaspwa}.

   Further special limits also arise.  For example, if we take
$p=q=r=1$, the roots $x_1$ and $x_2$ coalesce, $\alpha=\beta$, and
the metric becomes the homogeneous $T^{1,1}$ space, with the
four-dimensional base space being $S^2\times S^2$.  In another limit,
we can set $\mu=0$ in (\ref{d4met}) and obtain
the round metric on $S^5$, with $CP^2$ as the base. (In fact,
we obtain $S^5/Z_q$ if $p=0$.) Except in these
special ``regular'' cases, the four-dimensional base spaces themselves
are singular, even though the Einstein-Sasaki spaces $L^{p,q,r}$ are
non-singular.  The Einstein-Sasaki space is called quasi-regular if
$\del/\del\tau$ has closed orbits, which happens if $c_1$ is rational.
If $c_1$ is irrational the orbits of $\del/\del\tau$ never close, and
the Einstein-Sasaki space is called irregular.

   Our construction generalises straightforwardly to all odd higher
dimensions $D=2n+1$.  We take the rotating Kerr-de Sitter metrics
obtained in \cite{gilupapo1,gilupapo2}, and impose the Bogomol'nyi
conditions $E- g\sum_i J_i=0$, where $E$ and $J_i$ are the energy and
angular momenta that were calculated in \cite{gibperpop}, and given in
(\ref{ejrels}).  We find that a non-trivial BPS limit exists where $g
a_i = 1 - \ft12\alpha_i \epsilon$ and $m= m_0 \epsilon^{n+1}$.  After
Euclideanisation, we obtain $D=2n+1$ dimensional Einstein-Sasaki
metrics $ds^2$, given by
\be
\lambda ds^2 = (d\tau+\sigma)^2 + d\bar s^2\,,
\ee
with $R_{\mu\nu}=2n\lambda g_{\mu\nu}$, where the $2n$-dimensional
metric $d\bar s^2$ is Einstein-K\"ahler, with K\"ahler form
$J=\ft12d\sigma$, and
\bea
d\bar s^2 &=& \fft{Y dx^2}{4x F} - \fft{x(1-F)}{Y}
\Big(\sum_i \alpha_i^{-1}\,  \mu_i^2 d\varphi_i\Big)^2 \nn\\
&&+ \sum_i (1-\alpha_i^{-1}\, x)(d\mu_i^2 + \mu_i^2 d\varphi_i^2)
\nn\\
&&+ \fft{x}{\sum_i \alpha_i^{-1} \mu_i^2}\, 
        \Big( \sum_j \alpha_j^{-1}\, \mu_j d\mu_j\Big)^2 
-\sigma^2\,,\nn\\
\sigma &=& \sum_i (1-\alpha_i^{-1}x)\mu_i^2\, d\varphi_i\,,\\
Y&=&\sum_i\fft{\mu_i^2}{\alpha_i-x}\,,\qquad 
  F= 1- \fft{\mu}{x}\, \prod_i(\alpha_i-x)^{-1}\,,\nn
\eea
where $\sum_i \mu_i^2=1$.

    The discussion of the global properties is completely analogous to
the one we gave previously for the five-dimensional case.  The $n$
Killing vectors $\del/\del\varphi_i$ vanish at the degenerations of
the $U(1)^{n+1}$ principal orbits at $\mu_i=0$, and conical
singularities are avoided if each coordinate $\varphi_i$ has period
$2\pi$.  The Killing vectors
\be
\ell_i = c(i) \,\fft{\del}{\del\tau} + 
\sum_j a_j(i)\, \fft{\del}{\del\varphi_j} 
\ee
vanish at the roots $x=x_i$ of $F(x)$, and have unit surface gravities 
there, where
\be
a_j(i) = -\fft{c(i) \alpha_j}{\alpha_j-x_i}\,,\quad
c(i)^{-1} = \sum_j \fft{x_i}{\alpha_j-x_i} -1\,.
\ee
The metrics extend smoothly onto complete and non-singular manifolds
if $p \ell_1 + q\ell_2 + \sum_j r_j \del/\del\varphi_j=0$ for coprime
integers $(p,q,r_j)$, with coprimality conditions on $p$ and $q$ 
and the $r_i$. This implies the algebraic equations
\be
p c(1) + q c(2)=0\,,\quad
p a_j(1) + q a_j(2) + r_j=0\,,\label{pqrj}
\ee
determining the roots $x_1$ and $x_2$, and the parameters $\alpha_j$.
The two roots of $F(x)$ must be chosen so that $F>0$ when $x_1<x<x_2$.
With these conditions satisfied, we obtain infinitely many new
complete and non-singular Einstein-Sasaki spaces in all odd dimensions
$D=2n+1$.  Since it follows from (\ref{pqrj}) that $p+q=\sum_j r_j$,
these Einstein-Sasaki spaces, which we denote by $L^{p,q,r_1,\cdots
,r_{n-1}}$, are characterised by specifying $(n+1)$ coprime integers,
with coprimality conditions on $p$ and $q$ and the $r_i$,
which must lie in an appropriate domain.  The $n$-torus of the
$\varphi_j$ coordinates is in general factored by a freely-acting
$Z_k$, where $k=\hbox{gcd}\, (p,q)$.  The volume (with $\lambda=1$) is
given by
\be
V= \fft{|c(1)|}{q}\, {\cal A}_{2n+1}\, [\prod_i 
 \Big(1-\fft{x_1}{\alpha_i}\Big) -\prod_i\Big(1-\fft{x_2}{\alpha_i}\Big)]\,,
\ee
since $\Delta\tau$ is given by $2\pi k|c(1)|/q$, and ${\cal A}_{2n+1}$
is the volume of the unit $(2n+1)$-sphere.  In the special case that
the rotations $\alpha_i$ are set equal, the metrics reduce to those
obtained in \cite{gamaspwa2}.

   Finally, we note that we also obtain new complete and non-singular
Einstein spaces in $D=2n+1$ that are not Einstein-Sasaki, by taking
the Euclideanised Kerr-de Sitter metrics of \cite{gilupapo1,gilupapo2}
and applying the analogous criteria for non-singularity at degenerate
orbits that we have introduced in this paper.  Thus we Euclideanise
the metrics given in equation (3.5) of \cite{gilupapo1} by sending
$\tau\rightarrow -\im \tau$, $a_i\rightarrow \im\alpha_i$, take
Killing vectors $\ell_i$ that vanish on two adjacent horizons, and
have unit surface gravities, obtained from the $\ell$ given in
equation (4.7) of \cite{gilupapo1} by dividing by the surface gravity
in equation (4.17), and then impose the rationality conditions
following from $p \ell_1 + q \ell_2 + \sum_j r_j
\del/\del\varphi_j=0$.  This gives infinitely many new examples of
complete and non-singular Einstein spaces, beyond those obtained in
\cite{gilupapo1}.  They are characterised by $(n+2)$ coprime integers,
and we shall denote them by $K^{p,q,r_1,\ldots,r_n}$.

   Further details of these results will appear in \cite{cvlupapo3}.

\noindent{\bf Note added}: In a private communication, Krzysztof Galicki
has told us of a simple argument showing that all the $L^{p,q,r}$ spaces
are diffeomorphic to $S^2\times S^3$, since the total space of the 
Calabi-Yau cone can be viewed as a symplectic quotient of $C^4$ by
the diagonal action of $S^1(p,q,-r,-s)$ with $p+q-r-s=0$. 

\noindent{\bf Acknowledgements}: 

   We thank Peng Gao, Gary Gibbons, Joan Sim\'on and James Sparks for useful
discussions.  M.C. and D.N.P. are grateful to the George P. \& Cynthia
W. Mitchell Institute for Fundamental Physics for hospitality.
Research supported in part by DOE grants DE-FG02-95ER40893 and
DE-FG03-95ER40917, and the NSERC of Canada.

\end{document}